\documentclass[a4paper]{ESASPCS13Style}
\usepackage{epsfig}

\begin{document}

\title{The L  to T dwarf Transition}

\author{M.\,S. Marley\inst{1}, M.\,C. Cushing\inst{1}, \and D. Saumon\inst{2} } \institute{NASA Ames Research Center, Mail Stop 245-3, Moffett Field CA, USA
  \and Applied Physics Division, Los Alamos National Laboratory, MS F699, Los Alamos, NM, USA}

\maketitle 

\begin{abstract}

While the precise mechanism responsible for the L to T dwarf transition remains unclear, it is clearly caused by changing cloud characteristics.  Here we briefly review data relevant to understanding the nature of the transition and argue that changing atmospheric dynamics produce the transition by opening holes through the global iron and silicate cloud decks.  Other possibilities, such as a sudden vertical collapse in these cloud decks are also considered.
Any acceptable model of the L to T transition must ultimately connect changing cloud properties to the underlying atmospheric dynamics.

\end{abstract}

\section{Introduction}
  
 As the atmospheres of brown dwarfs cool with time, their spectral signatures
reflect a progression of changes in their atmospheric chemical equilibrium
and condensate structure.  In M dwarfs the elements O, C, and N are
predominantly found in $\rm H_2$O, CO, and $\rm N_2$ and the atmosphere is too
warm for condensation of solids (Allard \& Hauschildt 1995; Lodders 1999).  As the effective
temperature ($T_{\rm eff}$) falls, a variety of condensates form in the atmosphere, most notably
iron and silicates.  These condensates are apparently not well-mixed
through the atmosphere, but  are found in discrete
cloud layers overlying the condensation level (Ackerman \& Marley 2001; Marley et al. 2002; Tsuji 2002; Woitke \& Helling 2004). 

\begin{figure}[!t]
\centerline{\hbox{\includegraphics[height=4.5in,angle=0]{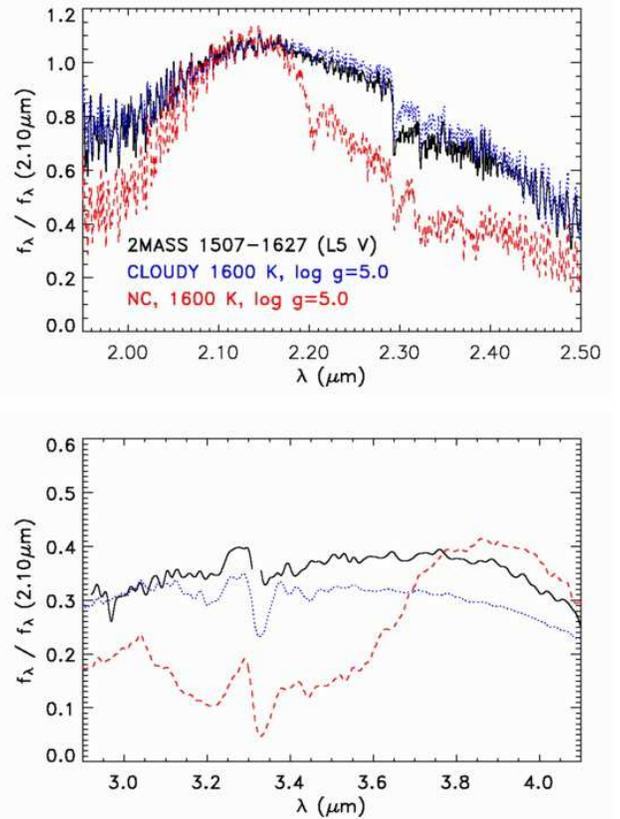}}}
\caption{Comparison of $K$ (top) and $L$-band (bottom) spectra (solid, black line) of the L5 dwarf 2MASS1507-1627 obtained by Cushing (2003) with SpeX at the IRTF to clear (long dashed, red curve in each panel) and cloudy (short-dashed, blue, $f_{\rm sed}=3$) models by Marley et al. (2003).  The best fit, by eye, spectrum is for a cloudy model with $T_{\rm eff}=$1600 K which is consistent with the measurement by Golimowski et al. (2004) of 1475 -- 1800 K based on the bolometric luminosity. Note that the signature of the $3.3\,\rm\mu m$-methane band is much stronger in the cloud free model since the atmosphere is cooler and the equilibrium abundance of $\rm CH_4$ is correspondingly larger.}
\end{figure}

By the time the  $T_{\rm eff}$ falls to that of a late L dwarf the cloud layer is optically thick and affects either directly (as
a major opacity source) or indirectly (by altering the atmospheric
temperature/pressure profile) all spectral regions.  The exact spectral
signature of the cloud depends both on its vertical thickness and the particle
size distribution of the condensates.  In addition, as the atmosphere cools,
chemical equilibrium begins to favor first $\rm CH_4$ over CO and then
$\rm NH_3$ over $\rm N_2$ (Tsuji 1964; Fegley \& Lodders 1996).  Thus $\rm CH_4$ absorption
in the $K$ band begins to replace CO and $\rm NH_3$ appears (Roellig et al. 2004) in the mid-infrared by the late L's.  By the early to mid T dwarfs the condensate cloud is forming quite deep in
the atmosphere.  In the relatively clear, cool atmosphere above the cloud,
chemical equilibrium begins to strongly favor $\rm CH_4$ and $\rm NH_3$ and
their spectral features, along with particularly strong bands of water, grow
in prominence (Marley et al. 1996, Burrows et al. 1997, Allard et al. 2001, Burrows et al. 2003).  Figures 1 through 3 illustrate these spectral trends in the $K$ and $L$ bands for L5 through T5 brown dwarfs.

\begin{figure}[!t]
\centerline{\hbox{\includegraphics[height=4.5in,angle=0]{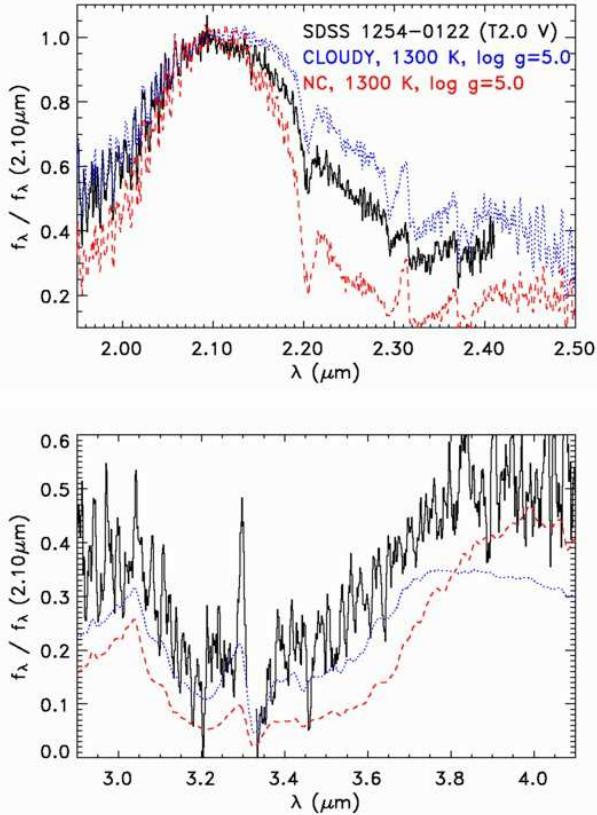}}}
\caption{Comparison of $K$ (top) and $L$-band (bottom) spectra of the T2 dwarf SDSS1254-0122 obtained by Cushing (2003) with SpeX at the IRTF to clear and cloudy ($f_{\rm sed}=3$) models by Marley et al. (2003) (line types as in Figure 1).  The best fit, by eye, models have $T_{\rm eff}=$1300 K, consistent  with the measurement by Golimowski et al. (2004) of 1150 -- 1500 K based on the bolometric luminosity. Note that the $K$ band spectral data lie between the clear, no-cloud and the cloudy models. }
\end{figure}
\begin{figure}[!t]
\centerline{\hbox{\includegraphics[height=4.5in,angle=0]{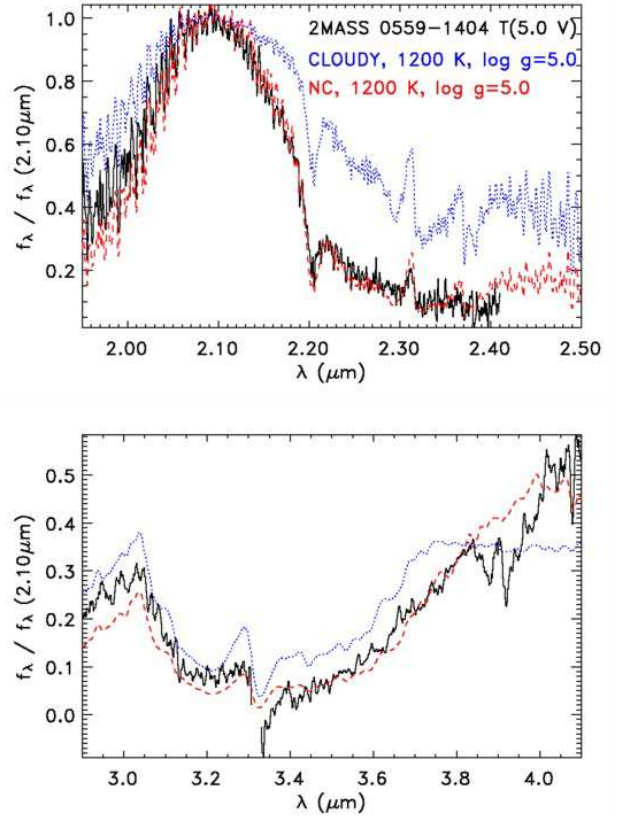}}}
\caption{Comparison of $K$ (top) and $L$-band (bottom) spectra of the T5 dwarf 2MASS0559-1404 obtained by Cushing (2003) with SpeX at the IRTF to clear and cloudy ($f_{\rm sed}=3$) models by Marley et al. (2003) (line types as in Figure 1).  The best fit, by eye, spectrum is for a clear model with $T_{\rm eff}=$1200 K, consistent with the measurement by Golimowski et al. (2004) of 1150 -- 1500 K based on the bolometric luminosity. The spectral feature at $3.93\,\rm \mu m$ in the data arises from incomplete removal of telluric $N_2O$ absorption.}
\end{figure}

\section{Signatures of the L to T Transition}
Below we summarize the characteristics exhibited by brown dwarfs at and near the L to T transition (approximately L8 to T5):

{\it Turn to the blue in $J-K$:}  The colors of L dwarfs become progressively redder until they saturate at $J-K \sim 2$ at spectral type L8 (Knapp et al. 2004).  This color then rapidly turns to the blue, reaching $J-K\sim -0.8$ by T8 or so. 

{\it Color change at near constant $T_{\rm eff}$:}  Recent estimates of the bolometric $T_{\rm eff}$ from Golimowski et al. (2004) have quantified the rapid rate of this color change, as shown in Figures 4 and 5.  Most ($>80\%$) of the change is $J-K$ color is seen to occur over a very small $T_{\rm eff}$ range near 1300 K.   This is a remarkable result as it implies that brown dwarfs are undergoing substantial spectral and color changes over a very small temperature range.

{\it Brightening at J Band:} The L to T transition also appears to be associated with a brightening at $J$ band from late L to early T (T4 or so) (Knapp et al. 2004).  $H$, $K$, $L$, and $M$ bands show no sign of such brightening (Knapp et al. 2004; Golimowski et al. 2004), while there is some evidence of a brightening at $Z$.  It should be noted that the bolometric luminosity, as would be expected, does {\it not} increase across the transition (Golimowski et al. 2004).

{\it Resurgence of FeH:} Burgasser et al. (2002) argue there is evidence that, after decaying away as FeH is presumably lost to Fe drops and grains,  the $0.997\,\rm\mu m$ FeH band shows a resurgence in strength, coincident with the $J-K$ color change.

{\it Model Spectral Fits:}  The comparison of models and data shown in Figures 1 through 3 provides additional information about the transition.   In Figure 1 a cloudy model does a good job of reproducing the $K$-band spectra of an L5 dwarf. A model with no cloud opacity predicts too much methane absorption in both $K$ and $L$ bands as well as a too-deep water band.  Comparing the cloudy and cloudless models for this objects makes clear
why the $J-K$ color is such an important diagnostic for the cloud. In the $L$ band the model gets the depth of the $3.3\,\rm\mu m$ methane band correct, which suggests the thermal structure of the model and the associated equilibrium methane abundance are reasonable.  

By T2 (Figure 2), however, a model using the same cloud model (Ackerman \& Marley 2001) is apparently somewhat too warm, predicting a bit too much CO and too little $\rm CH_4$.  At $K$ band the observed spectrum lies between this cloudy model and a cloudless model.  The overall shape of the $L$ band spectrum, which probes higher in the atmosphere, seems to be best fit by  a combination of the cloudy and cloudless models.  Interestingly the amplitude of the methane feature at $3.3\,\rm \mu m$ is larger than either the cloudy or cloudless models predict, which may indicate that the temperature gradient in the photosphere above the cloud deck is steeper than either model predicts.

 Finally by T5 (Figure 3) a model with no cloud opacity (but with condensation included in the equilibrium chemistry) fits very well both at $K$ and $L$ bands, implying that condensates play a very small role in controlling the thermal profile and emitted flux.  The difference between the best fitting models for the T2 and the T5 dwarfs is only $100\,\rm K$!

\begin{figure}[t]
\centerline{\hbox{\includegraphics[height=3in,angle=0]{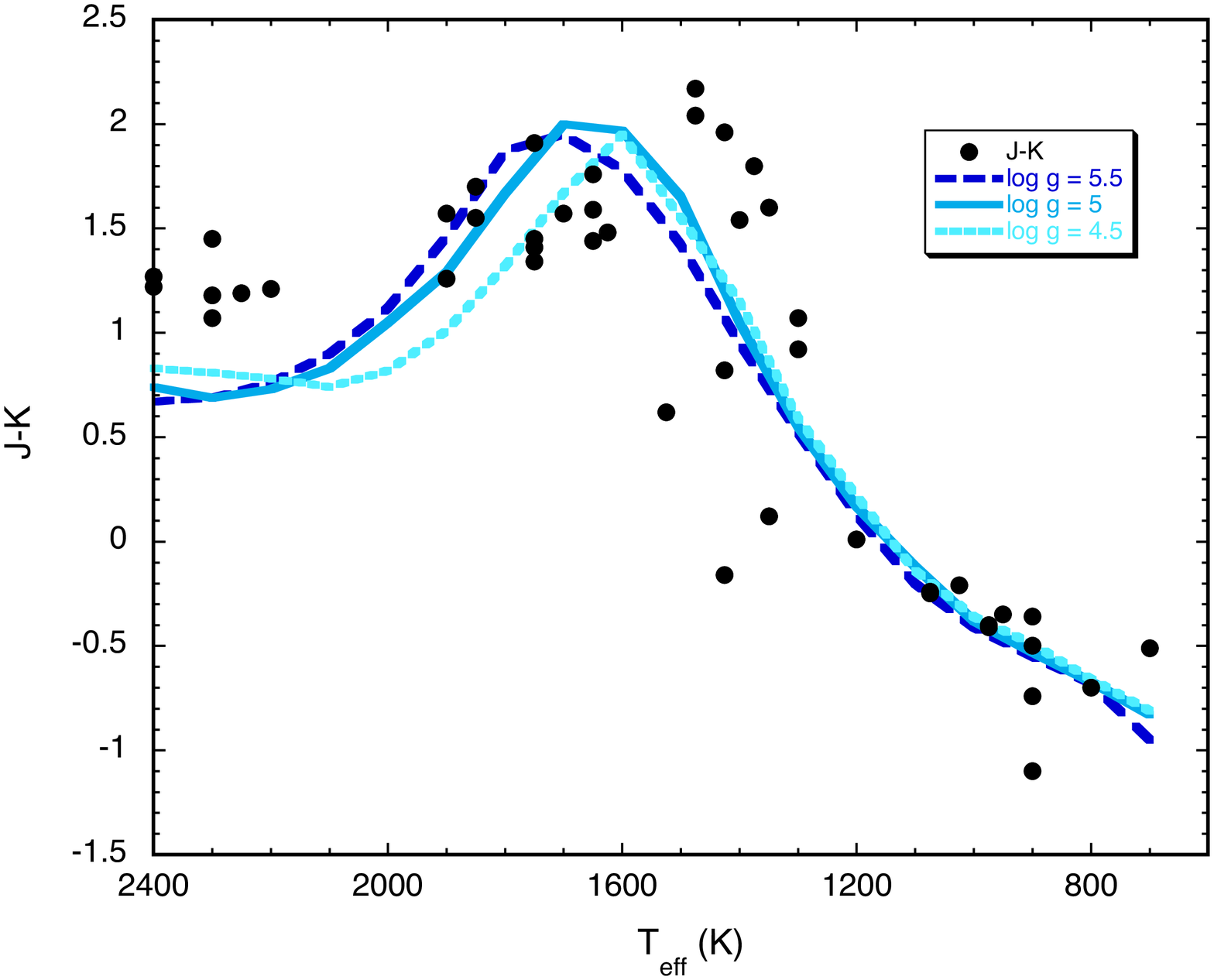}}}
\caption{Points show ultracool dwarf MKO $J-K$ color and $T_{\rm eff}$ from Knapp et al. (2004) and Golimowski et al. (2004).
Curves show $J-K$ predicted for three different $\log g$ by Unified Cloudy Models kindly provided by T. Tsuji (Tsuji et al. 2004).  Note that even accounting for the likely spread in $\log g$, the UCM models do not exhibit the sharp blueward turn at near constant $T_{\rm eff}$ as seen in the data. }
\end{figure}

\begin{figure}[t]
\centerline{\hbox{\includegraphics[height=3in,angle=0]{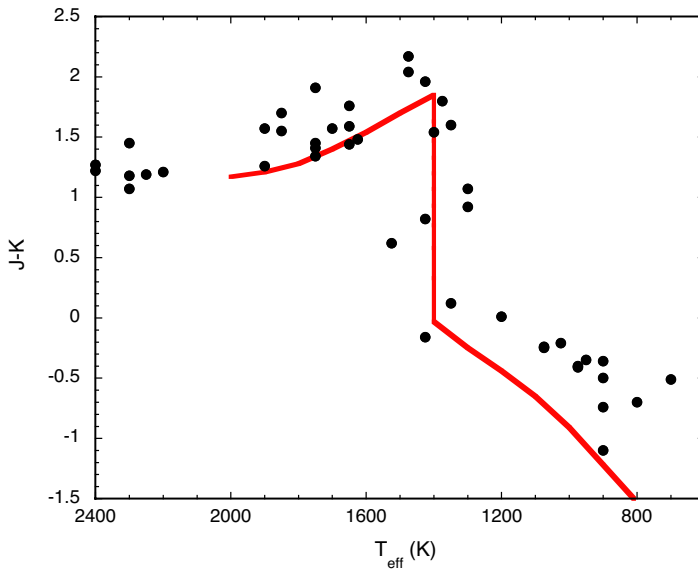}}}
\caption{Data points same as in Figure 4.  Model curve is a composite of cloudy models (for $T_{\rm eff} \ge 1400\,\rm K$)
and clear (for  $T_{\rm eff} \le 1400\,\rm K$) models (Marley et al. 2002) connected by a vertical line at $T_{\rm eff}=1400\,\rm K$, all for $\log g = 5$.} 

\end{figure}

\section{The Transition Mechanism}
Any explanation of the L to T transition mechanism must be consistent with the evidence summarized above. The unmistakable gross explanation--that condensates have been lost from the atmosphere--belies the difficulty in explaining this loss is a self-consistent manner.  That a sinking, finite-thickness cloud deck will eventually disappear from sight allowing the atmosphere above to cool has been apparent for some time (Marley 2000, Allard et al. 2001, Marley et al. 2002, Tsuji 2002).  The difficulty lies in explaining the rapidity of the color change in light of the measured effective temperatures (Figures 4 and 5). For example while nicely accounting for the $J-K$ colors of the reddest L dwarfs, the model of Ackerman \& Marley (2001)  takes much too long to ultimately sink out of sight (Burgasser et al. 2002; Knapp et al. 2004).  

In a series of papers Tsuji (Tsuji 2001, Tsuji \& Nakajima 2003, Tsuji et al. 2004) proposed that a physically thin cloud, thinner than predicted by the Ackerman \& Marley model, could self-consistently explain the rapid L to T transition.  These `UCM' models indeed exhibit a faster L- to T-like transition, but as Figure 4 demonstrates the UCM models are not consistent with the observed rapidity of the color change.  Even accounting for a likely spread in gravities across the transition can not account for the observations. In addition the UCM models, like the cloudy models of Marley et al., do not brighten in $J$ band across the transition.  Tsuji et al. had to invoke an exceptionally large spread in atmospheric $\log g$ of known sources across the transition in order to account for both the reddest and dimmest late L dwarfs and the brightest and bluest early T dwarfs.  Finally the UCM models could not explain the resurgence in FeH that is observed across the transition.    

To overcome the sort of difficulties faced by the Tsuji et al. models, Burgasser et al. (2002), following a suggestion from
Ackerman \& Marley (2001),  hypothesized that at the L to T transition the global cloud deck rapidly breaks apart.  Under this scenario holes in the cloud deck appear at $T_{\rm eff} \sim 1300$ to $1400\,\rm K$.  In the molecular window regions, particularly $Z$ and $J$ bands, bright flux from deeply seated regions pours out of the holes left by the departure of the cloud deck.  This outpouring of flux is then responsible for the rapid color change in $J-K$ (Figure 5), the brightening in $J$ (and apparently also $Z$) band, and the reappearance of FeH.   The fact that the T2 dwarf (Figure 2) seems to be a composite of the cloud free and cloudy model spectra supports this interpretation.  

Burgasser et al. suggest that the cloud holes appear when the combined iron and silicate clouds sink sufficiently deeply into the global convection zone.  On Earth clouds tend to be more spatially uniform when they form in relatively shallow convective layers.  Regions in which the convective layer is thick, such as near the equator, seem to be inhabited by towering cumulus clouds separated by cloud-free regions.  The presence of some relatively cloud free regions in the atmospheres of Venus and Jupiter provides evidence that cloud layers are generally not globally uniform in planetary atmospheres and supports the plausibility of the mechanism.

Finally Knapp et al. (2004) proposed a third alternative in which the sedimentation efficiency of the cloud substantially increases at the transition.  In this case the cloud remains homogeneous across the disk, but particle growth becomes much more efficient.  Efficient growth leads to larger particles which more rapidly fall out of the atmosphere.  This leads to optically thinner clouds.  In the language of Ackerman \& Marley (2001) this is described as $f_{\rm sed}\rightarrow \infty$.  As discussed elsewhere in these proceedings, Tsuji and collaborators now favor a sudden collapse of the global cloud deck ($T_{\rm crit} \rightarrow T_{\rm cond}$) at the transition.  This is similar to the Knapp et al. (2004) suggestion with the exception that Tsuji et al. do not address the particle size.  

Regardless of whether the L to T transition is explained by the appearance of holes in the global cloud deck or a sudden increase in the efficiency of condensate sedimentation, the root cause must lie with the atmospheric dynamics.  What aspect of atmospheric circulation or dynamics would favor the appearance of holes or the sudden collapse of the cloud deck?  Perhaps the behavior of condensates change when the cloud reaches a certain depth in the atmospheric convection zone or perhaps the second, detached convection zone found in brown dwarf atmosphere models (Marley et al. 1996, Burrows et al. 1997, Allard et al. 2001, Tsuji 2002) plays a role.  Another possibility is that there is a change in the global atmospheric circulation that affects the behavior of the cloud decks.  Schubert \& Zhang (2000) found that brown dwarfs likely exhibit one of two styles of global atmospheric circulation: dominated by rotation, like Jupiter, or fairly independent of rotation, like the sun.    Since the luminosity falls with age, the Rayleigh and Eckman numbers of brown dwarfs,
which influence the regime in which the atmospheric dynamics falls, likewise vary with time.  Perhaps the L to T transition is associated with a change between the two regimes.
Until such possible mechanisms have been quantitatively addressed the nature of the L to T transition will remain the domain of plausible, if ad hoc, modeling.

\begin{acknowledgements}
MSM acknowledges support from NASA grants NAG2-6007 and NAG5-8919 and NSF grant AST 00-86288.  MC acknowledges support from the Spitzer Fellow Program. This work was supported in part by the U.S. Department of Energy under contract W-7405-ENG-36.

\end{acknowledgements}

\end{document}